\documentclass[journal]{IEEEtran}
\usepackage{amsmath}
\usepackage{url}
\usepackage{enumerate}
\usepackage{graphicx}
\usepackage{color}
\usepackage{paralist}

\usepackage{hyperref}
\hypersetup{%
plainpages=false,
colorlinks,urlcolor=blue,linkcolor=blue,citecolor=blue,
pdftitle={Using extreme value theory to smooth demand-net-of-wind when
  estimating the loss of load expectation and expected energy unserved
  for capacity adequacy assessment},
pdfauthor={Amy Wilson and Stan Zachary},
pdfsubject={},
pdfpagemode={UseOutlines},
pdfpagelayout={OneColumn},
pdfstartview={Dubhe}
}

\begin{document}

\title{Using extreme value theory for the estimation of risk metrics
  for capacity adequacy assessment}

\author{Amy L.~Wilson 
        and Stan~Zachary% ,

        \thanks{The authors acknowledge support from EPSRC grants
          EP/N030028/1 and EP/K036211/1 as well as from National Grid
          and the School of Mathematics at the University of
          Edinburgh.}% <-this % stops a space

        \thanks{A.L. Wilson and S. Zachary are with the School of
          Mathematics, University of Edinburgh, Edinburgh, EH9 3FD,
          Scotland, UK (email: Amy.L.Wilson@ed.ac.uk,
          s.zachary@gmail.com).}}%

\maketitle

\begin{abstract}
  This paper investigates the use of extreme value theory for
  modelling the distribution of demand-net-of-wind for capacity
  adequacy assessment.  Extreme value theory approaches are
  well-established and mathematically justified methods for estimating
  the tails of a distribution and so are ideally suited for problems
  in capacity adequacy, where normally only the tails of the relevant
  distributions are significant.  The extreme value theory peaks over
  threshold approach is applied directly to observations of
  demand-net-of-wind, meaning that no assumption is needed about the
  nature of any dependence between demand and wind.

  The methodology is tested on data from Great Britain and compared to
  two alternative approaches: use of the empirical distribution of
  demand-net-of-wind and use of a model which assumes independence
  between demand and wind.  Extreme value theory is shown to produce
  broadly similar estimates of risk metrics to the use of the above
  empirical distribution but with smaller sampling uncertainty.
  Estimates of risk metrics differ when the approach assuming
  independence is used, especially when data across different
  historical years are pooled, suggesting that assuming independence
  might result in the over- or under-estimation of risk metrics.

\end{abstract}


\section{Introduction}
\label{intro}

Wind generation plays an increasingly important role in the global
supply of electricity.  However, in the context of capacity assessment
or security-of-supply the contribution of this wind generation can be
difficult to quantify (see \cite{Keane2011} and \cite{Milligan2016}
for a survey of current approaches).  This difficulty arises because,
for capacity assessment, what matters most is the contribution of wind
at the times of very peak demand when the system is typically under
most stress.  For example, in Great Britain (GB) this peak demand
usually occurs in the early evening in winter when the weather is
extremely cold.  Because peak demand events are rare, and may scarcely
occur at all in years with milder weather, there is relatively little
data with which to make an accurate assessment of what the wind is
doing at such times.  Moreover, demand patterns are known to change
through time, limiting the number of years of data suitable for
estimating the relevant demand-wind relationship.

A particular concern is that the cold winter weather associated with
the highest electricity demands may be associated with large-scale
weather systems that lead to low wind conditions. There is some debate
in the literature on this issue (see e.g.\ \cite{Brayshaw2012} and
\cite{Thornton2017} and the references therein) but there is certainly
insufficient evidence to suggest that there is no association at all
between demand and wind (either positive or negative correlation). A
failure to account for any reduction in wind at times of high demand
would lead to overestimation of the contribution of wind to capacity
adequacy.

The objective of a capacity adequacy study is to assess the risk of
insufficient electricity generation to meet demand for some future
year or season of interest.  (Here by ``season'' is meant the peak
demand season within each year, for example the winter months within
GB.)  Typically this is achieved using a risk metric such as
\emph{loss of load expectation} (LoLE) or \emph{expected energy
  unserved} (EEU) (\cite{NERC2018}, \cite{Amelin2009}).  Such expected
value risk metrics may be completely defined in terms of a
\emph{non-sequential} or \emph{snapshot} model which integrates over
the course of the future season the joint sequential distribution of
relevant variables such as demand, wind generation and conventional
generation---without regard to the temporal structure of this joint
distribution within the season (which is not necessary for the
definition of such expected value risk metrics).  Alternatively a
non-sequential model may be viewed as that defining the distribution
of the above variables at a uniformly randomly sampled point in time
during the season under study (see \cite{Zachary2012} for further
details).

The basic non-sequential probability model is well established, and
used in Great Britain, the USA, and elsewhere (e.g.~\cite{NG2016},
\cite{NERC2018}).  The model consists of a specification of the joint
distribution as described above of the random variables $X$, $W$ and
$D$ which represent respectively \emph{conventional generation},
\emph{wind generation} and \emph{demand}.  Then the random variable
\begin{equation}
  Z = X+W-D \label{sdbalance}
\end{equation}
models the corresponding excess of supply over demand.  The model
\eqref{sdbalance} has two major components:
\begin{enumerate}[(a)]
\item the (non-sequential) distribution of \emph{demand-net-of-wind}
  $D-W$, which requires estimation from data;
\item the (non-sequential) distribution of \emph{conventional
    generation} $X$, which is usually given by a fully specified
  probabilistic model.
\end{enumerate}
The variables $D-W$ and $X$ are assumed probabilistically independent,
so that the distribution of their difference $Z$ is obtained by
convolution.  It is this distribution of \emph{the supply-demand
  balance}~$Z$ which is the primary output of the non-sequential
model, and from which the above risk metrics LoLE and EEU, and other
statistics of interest, are calculated.  For more details of the
underlying model see \cite{Wilson2016}, \cite{Zachary2012},
\cite{Dent2014}, \cite{Dent2016}.  In particular we have
% the loss of load expectation (LoLE) and expected energy unserved (EEU) by
\begin{align}
\mbox{LoLE} =&nP(Z < 0), \label{LoLE}\\
\mbox{EEU}=&nE(\max(-Z, 0))=n\int^0_{-\infty} \mathrm{d}z P(Z<z) \label{EEU}
\end{align}
where $n$ is the number of hours in the future season under study.

A standard approach for modelling the non-sequential distribution of
conventional generation $X$ is to use independent two-state models for
individual generators and then to use convolution to obtain the
distribution of the total available generation \cite{Billinton2012}.
This paper is instead primarily concerned with the estimation of the
distribution of demand-net-of-wind $D-W$.  For the `future' season to
be studied, this distribution is typically estimated from a dataset of
hourly historical paired observations of (demand, wind speed) made in
earlier seasons.  Each such historical season of demand observations
is `forward-mapped' by re-scaling to the future season under study.
Forward-mapped wind generation observations for the future season are
obtained by combining the historical wind speed measurements over a
geographic grid with a physical model for the capacities, locations
and power curves of the installed wind generation in the future
season.  For each historical season in the dataset this process yields
a forward-mapped hourly trace $(d_t, w_t)$ for
$t \in \{1, \ldots, n\}$ of aggregate demand and wind generation for
the future season under study.  Estimates of risk metrics can be based
on data from individual historic seasons or based on data pooled over
multiple historic seasons.

One approach for estimating the distribution of $D-W$ is simply to use
the empirical distribution of the forward-mapped observations
$d_t-w_t$ (this is sometimes known as {\it hindcast})
(\cite{Zachary2012}, \cite{Keane2011}).  This approach makes no
assumption about the relationship between wind generation and demand.
However, as we discuss above and illustrate in Section~\ref{Data},
there is typically very little data available for estimating the part
of the distribution of $D-W$ that is relevant for capacity adequacy,
i.e.\ the far right tail.  Further, as illustrated in
Figure~\ref{ScatterContours}, the presence or absence of a single
observation may, in the hindcast approach, have a very considerable
influence on the estimated values of risk metrics such as LoLE and
EEU.  There is thus considerable concern as to the reliability of such
estimates obtained using the hindcast approach.

An alternative approach to the estimation of the distribution of $D-W$
is to estimate the distribution of demand~$D$ and also the
distribution of wind~$W$ conditional on demand~$D$.  The simplest
possibility here is to assume that demand and wind are
\emph{independent}, in which case the distribution of wind may be
estimated using all wind observations, not just those relatively few
corresponding to times of high demand.
% The advantage of this method is that more data can be used to
% estimate the distribution of wind.
However, as described above, there is concern that in GB, for example,
the assumption of independence may not hold in practice.  A more
general approach is to develop a parametric (or smoothed) model for
wind generation conditional on demand---see, e.g.~\cite{Wilson2018b}.
% follows this approach but supplements observations of demand with
% those of the variable temperature (closely associated with demand in
% GB) and makes use of the fact that there is considerably more
% historical data with which to estimate the temperature-wind
% relationship; however this approach requires quite strong
% assumptions to be made about the nature of the dependence between
% demand, wind and temperature, and these are difficult to fully
% validate.  develops a parametric model for wind generation
% conditional on temperature to approximate the distribution of wind
% generation conditional on demand.  Again, the advantage is that more
% data can be used to estimate the distribution of wind because more
% paired (wind, temperature) data is available to fit the model than
% paired (wind, demand) data.  The disadvantage is that this method
% requires strong assumptions to be made about the nature of the
% dependence between demand, wind and temperature.

This paper proposes a method for estimating the distribution of $D-W$
using statistical extreme value theory (EVT) \cite{Coles2001}.  EVT is
a well-established methodology for making inference about the extremes
of distributions and is therefore well-suited to problems in capacity
adequacy, where, as discussed above and further in Section~\ref{Data},
interest is in the extreme right tail of the distribution of $D-W$.
EVT is based on asymptotic theory for the tails of distributions which
permits appropriate smoothing and, if necessary, extrapolation of
empirical data.  As with the hindcast approach, EVT uses directly the
empirical observations $d_t-w_t$ of $D-W$, without the need for any
assumptions about the statistical relationship between demand $D$ and
wind generation $W$.  The advantage of EVT is that smoothing the
empirical data reduces the influence of the very small number of
observations at the extremes of high demand and low wind.  If
appropriate, information about the shape of the far right tail of the
distribution of $D-W$ can be inferred from observations that are close
to the tail but not as extreme.  A further advantage of EVT is that it
does not make the assumption implicit in hindcast that there is no
possibility of the demand-net-of-wind in the future year or season
under study being more extreme than that observed historically.  In
\cite{Gao2018}, EVT is also used for capacity adequacy assessment, but
in combination with quantile regression models that incorporate
seasonal time effects.  The advantage of the present approach is that
it focuses on the use of EVT within the non-sequential model (which,
as remarked above, is sufficient for the risk metrics considered) and
avoids the considerable complications and distortions that arise in
fully accounting for seasonal effects--which arise on multiple time
scales (e.g.\ daily, weekly, yearly).  

The present methodology was developed in response to concerns from the
GB transmission system operator (National Grid) and was used to refine
the estimates of risk metrics in the GB capacity adequacy study (see
\cite{NG2016}). The GB system is therefore used as an exemplar.  While
the GB system is winter-peaking, this is not a necessary condition for
the methodology to be appropriate---the same techniques are applicable
for summer-peaking systems.

The rest of this paper is divided as follows: Section \ref{Data}
describes the data used for the analysis of the GB system, Section
\ref{Model} describes how EVT should be applied to model the supply
demand balance $Z$, and Section \ref{Results} presents and discusses
results for the GB system.

\section{Data}
\label{Data}

% If risk metrics are to be estimated separately for each historic
% year, $n$ is the number of hours in the future season under study.
% If risk metrics are to be estimated from multiple years of
% historical data pooled together, $n$ is the number of hours in the
% future season under study, multiplied by the number of available
% historical years.

As described in the Introduction, the distribution of $D-W$ is
estimated from hourly forward-mapped observations $(d_t, w_t)$,
$t \in \{1, \ldots, n\}$ of $D$ and $W$.  The season under study is
the winter season of 2014--15, where winter is defined as the 21 weeks
from the last Sunday in October.  In GB, the risk of a shortfall at
other times of year is negligible.  The data described in
\cite{Staffell2016} and \cite{Staffell2017} are used for the analysis.
These data are also used in \cite{Wilson2018b} and consist of:
\begin{compactitem}[-]
\item Hourly historical measurements of aggregate GB demand for the
  seven winter seasons from 2007--08 to 2013--14.\footnote{In GB,
    small (embedded) generators are not required to report their
    output to the transmission system operator so these demand
    measurements consist of the demand measured by the transmission
    system operator plus an estimate of the output of the embedded
    generators.} %
  Measurements for earlier seasons are available but these were not
  thought to be representative of current demand patterns. These
  aggregate demand measurements are forward-mapped to the 2014--15
  winter season by multiplying by an appropriate re-scaling factor for
  each of the seven seasons of historical data.  These re-scaling
  factors were determined by calculating the 90\% quantile of the
  daily maximum demands in each of the winter seasons from 1991--92 to
  2013--14.  A Lowess curve \cite{Cleveland79} was fitted to these
  90\% quantiles to obtain a smoothed curve estimating the variation
  in underlying demand over time.  The re-scaling factor for the
  $x$-th season was then set to the ratio of the fitted value of the
  Lowess curve for the 2013--14 season (see below) to the fitted value
  for the $x$-th season.

  This method of re-scaling historical demands by multiplying by some
  re-scaling factor is used in the GB capacity assessment study, where
  the re-scaling factor (calculated using a different methodology to
  that described above) is known as the Average Cold Spell (ACS) peak
  \cite{ACSpeak}.  By re-scaling, the aim is to adjust historical
  demands for general trends, such as those due to changes in the
  economy, but to preserve any variation between (winter) seasons due
  to changes in weather.  A Lowess curve is therefore appropriate
  because it smooths out any year-to-year fluctuations caused by
  random weather effects while still capturing long term trends in the
  re-scaling factor.  By fitting the Lowess curve to the 90\%
  quantiles rather than to the means the re-scaling is focused on the
  times of high demand that are of most interest in capacity
  assessment.

  Note that the historical demands are re-scaled to the 2013--14
  season, but the `future' season under study is 2014--15.  This
  mismatch is a result of data availability---the wind data related to
  January 2015, as described below, but the latest full year of demand
  data was 2013--14.  We have therefore made an assumption that demand
  conditions in 2013--14 are similar to those in 2014--15.  As the
  objective is to investigate methodology for assessing risk metrics
  this assumption has no effect on the conclusions drawn.
\item Hourly aggregate GB wind generation `observations' for the seven
  winter seasons from 2007--08 to 2013--14 forward-mapped to the
  2014--15 winter season to pair with the forward-mapped demand
  observations described above.  The wind generation observations were
  obtained by combining historical wind speed measurements (at the
  midpoint of each hour) with a model for the locations, capacities
  and power curves of the installed wind generation (approximately 14
  GW of installed capacity) in January 2015.  Aggregate wind
  generation for GB in each hour is then given by the sum of the wind
  power generated over all locations.
  % Wind capacity factors are then
  % calculated as the wind generation output divided by the total
  % installed capacity.  The hourly wind capacity factor represents the
  % instantaneous capacity factor mid-way through each hour.  It is these
  % historic wind capacity factors that were available for this
  % study. To scale the capacity factors to wind generation we
  % multiplied by 14GW, approximately representing the installed GB wind
  % generation in 2015.
\end{compactitem}

\begin{figure}
  \centering
  \includegraphics[width=\linewidth]{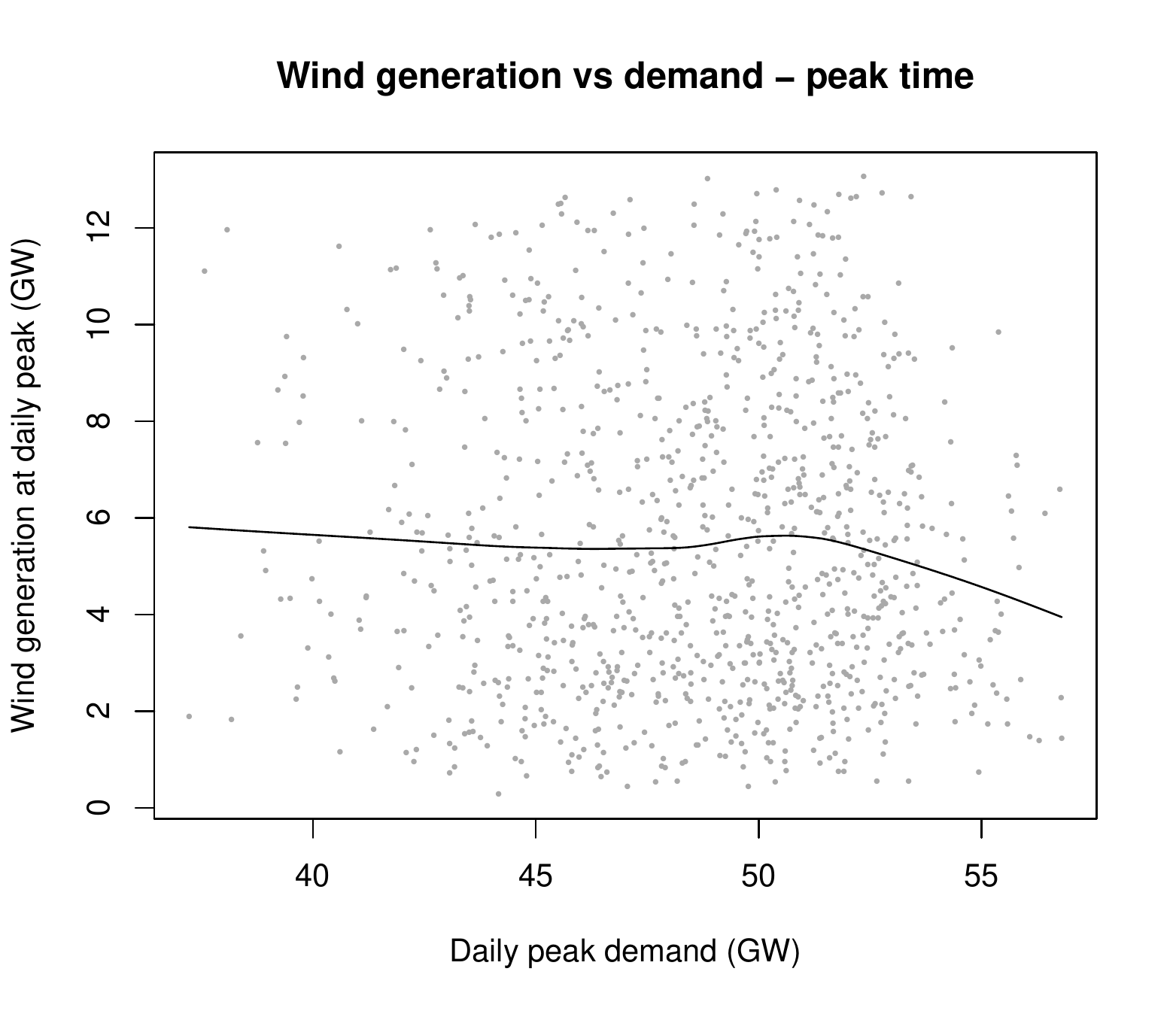}
  \caption{Wind generation against demand at time of daily peak demand
    for seven winter seasons in GB.  Overlaid is a Lowess curve
    showing the smoothed relationship between wind and
    demand.
    \label{ScatterLowess}}
\end{figure}

Figure~\ref{ScatterLowess} plots observations~$w_t$ of wind generation
against corresponding demand~$d_t$ at the times of daily peak demand
during the seven winter seasons comprising the dataset.  A smoothed
Lowess regression curve provides some evidence that the very highest
demands may be associated with lower wind generation.  Given the lack
of data in the extreme region of interest, there is insufficient
statistical evidence to decide the matter conclusively.  However, the
data do not justify any assumption of demand-wind independence.

\begin{figure}
  \centering
  \includegraphics[width=\linewidth]{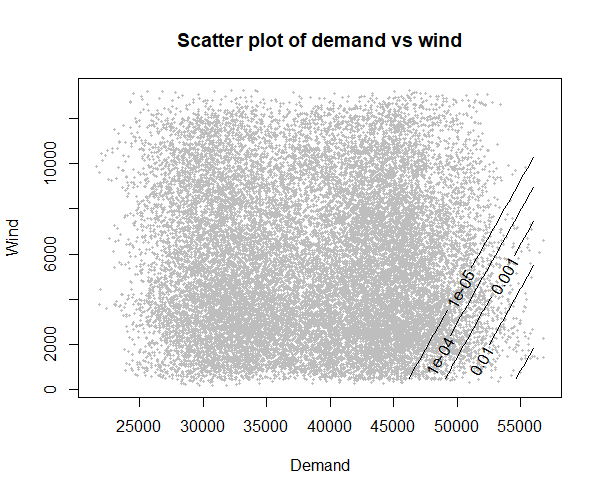}
  \caption{Hourly wind generation (MW) against demand (MW) for seven
    winter seasons in GB.  Contours show the contribution to LoLE of
    all points along the line.
    \label{ScatterContours}}
\end{figure}

Figure~\ref{ScatterContours} plots all the hourly forward-mapped
(demand, wind) data $(d_t,w_t)$ for the seven winter seasons
comprising the dataset.  The contour lines separate points according
to their values of $d_t-w_t$ and are such that the total contribution
to the LoLE, of a point along the line in a hindcast calculation (and
with the distribution of conventional generation~$X$ as described in
Section~\ref{estimation-lole-eeu}) would be as indicated.  Observe
that the only points that make a significant contribution to this risk
metric are indeed the very small number of observations in the lower
right corner, i.e.\ in the extreme right tail of the distribution of
$d_t-w_t$.

%\begin{figure}
%\centering
%\includegraphics[width=\linewidth]{Density_cont.pdf}
%\caption{Hourly wind generation (MW) against demand (MW) for seven
% winter seasons in GB.  Contours show the contribution to LoLE of all
% points along the line.  \label{DensityCont}}
%\end{figure}

\section{Methodology}
\label{Model}

\subsection{Statistical model}

We develop a statistical model for the marginal (non-sequential)
distribution of demand-net-of-wind ($D-W$) over the future season
under study.  The required result from EVT is that under appropriate
conditions, which we discuss below, the tail of the distribution of
$D-W$ is well-modelled by a generalised Pareto distribution (GPD).
Specifically, the excesses
\begin{equation}
  Y = D-W-u
\end{equation}
of $D-W$ above any sufficiently large threshold $u$, conditional on
$D-W>u$, have a distribution function $H$ given approximately by
\begin{equation}
  H(y) = P(Y \le y) = 1-\left(1+ \frac{\xi y}{\sigma_u} \right)^{-1/\xi},
  \label{eq1}
\end{equation}
for $y$ such that $y>0$ and $1+ \xi y/\sigma_u > 0$ (see Chapter 4 of
\cite{Coles2001}). Here the shape parameter $\xi$ is independent of
the threshold $u$ (for all $u$ sufficiently large that the
approximation \eqref{eq1} holds) and may be positive or negative,
corresponding to whether the distribution of $D-W$ is heavy- or
light-tailed.  The parameter $\sigma_u$, which may be thought of as a
scale parameter, depends on the threshold choice $u$, and increases
linearly with it at rate $\xi$.  The case $\xi = 0$ corresponds to
$D-W$ having an exponential tail.

Once an appropriate threshold $u$ is determined, and the parameters
$\xi$ and $\sigma_u$ of the GPD estimated, the full distribution of
demand-net-of-wind $D-W$ is given by its tail function
{\small
\begin{align}
P(D-W > v) =&P(D-W > u)P(D-W > v \mid D-W > u) \nonumber\\
  = & P(D-W > u)(1- H(v-u)), \ \  v\geq u, \label{eq2}
\end{align}}%
for values of $D-W$ (here denoted by $v$) in excess of the threshold
$u$.  Here the probability $P(D-W > u)$ that $D-W$ exceeds the chosen
threshold $u$ is is taken to be its empirically observed estimate,
i.e.\ the fraction of the observations of $D-W$ which exceed the
threshold $u$.  For values $v$ of $D-W$ below the threshold $u$ the
probability $P(D-W>v)$ is again taken to be its empirically observed
estimate.

Thus, for values of~$v$ below the threshold $u$, the estimates of the
probability $P(D-W>v)$ will be the same for the EVT and the hindcast
approaches, namely the empirically observed fraction of observations
exceeding $v$.  Where the use of EVT differs from the hindcast
approach is that, for $v$ above the threshold $u$, the empirical
estimate of $P(D-W>v)$ is replaced by the smooth function given by
\eqref{eq1} and \eqref{eq2}.  The effect of this smoothing is to
reduce the influence of the very small number of observations in the
extreme tail.  This is because, for large $v$ (perhaps considerably
greater than the threshold~$u$) the probability $P(D-W>v)$ is
estimated by suitably smoothing the distribution of all the
observations in excess of the threshold~$u$, and not simply by the
very small proportion of observations which may actually exceed~$v$.
(Clearly, for very large~$v$, the empirical estimate is very sensitive
to the precise number of observations in excess of~$v$.)  In
particular, the EVT approach allows us to estimate $P(D-W>v)$ for
values of $v$ in excess of the largest observed value $d_t-w_t$ of
$D-W$.

The result given in \eqref{eq1} is an asymptotic one.  It assumes that
the process of demand-net-of-wind has a sufficient degree of long-run
stationarity---despite the existence of shorter-run seasonal
variations---for the marginal distribution (represented above by the
random variable $D-W$) to be meaningful.  In addition there is an
assumption of some mild regularity conditions and an absence of
long-range dependence.  (Further details are given in
\cite{Coles2001}.)
% Note that although the
% demand-net-of-wind process itself is non-stationary (due to seasonal
% effects), the requirement is that the marginal distribution of
% demand-net-of-wind is stationary.
As historic years of data have been rescaled to the future year under
study, we expect these assumptions to be reasonable.

% The requirement that the marginal distribution of the process should
% be well-defined essentially translates to the idea that it should be
% statistically meaningful.  In the present context this is the act of
% faith which in general underlies capacity assessment methodology,
% namely that, with the usual adjustments for increasing overall
% levels of demand, historical years of demand and wind speed data
% should be representative of future years.  If this is accepted then
% the asymptotic theory is something which is best regarded as giving
% at least be a reasonable approximation.  In particular
The quality of fit by a GPD nevertheless requires empirical testing.
Empirical methods are also required in the selection of an appropriate
threshold $u$---one possibility is the examination of robustness of
estimated parameter values across a range of thresholds.  As we
demonstrate in Section~\ref{Results} the GPD fit in general works very
well for the GB data, and the parameter values are indeed robust with
respect to threshold variation within a reasonable range.

Note that although (demand, wind) is a bivariate process
our ultimate interest is in the univariate demand-net-of-wind
distribution.  The use of multivariate EVT methods to model the
bivariate distribution offers no further advantage here.

\subsection{Uncertainty}
\label{Modelb}

The aim of a capacity adequacy study is to assess the risk of
insufficient generating capacity to meet demand in the future season
under study.  The LoLE and EEU are expected value metrics in that they
give the long-run expected values of loss of load and energy unserved
respectively.  However, both wind generation and demand are very
dependent on the weather, which varies considerably from one (winter)
season to the next.  Thus, estimates of LoLE and EEU conditional on
the weather in a given season also vary considerably.  To fully
understand the risks to the system it is therefore important to
understand this weather-dependent variation.  This issue has become
more important as the proportion of total energy supplied by variable
generation has increased, as there has been a corresponding increase
in variation in loss of load duration and energy unserved from winter
season to winter season~\cite{Sheehy2016}.  This year-to-year
variation is reflected in the variation in the estimates of LoLE and
EEU based on the forward-mapped demand-net-of-wind traces associated
with individual historical seasons in our dataset.  We therefore
calculate these estimates based on individual historical seasons.  The
long-run LoLE and EEU are then estimated as the means of the
respective individual season-based estimates.
% The estimates conditional on individual historical years permit
% estimation of the variation in LoLE and EEU year-to-year due to
% weather effects.

It is also necessary to understand the statistical sampling
uncertainty associated with the long-run LoLE and EEU estimates. This
arises because there is considerable variation in weather conditions
between years, and the long run estimates are based on a sample of a
finite number (seven) of years of data. In addition to the
considerable variation observed above in the estimates of LoLE and EEU
based on the individual historical seasons of (demand, wind) data,
there are further, within each historical season, complex patterns of
dependence in the hourly forward-mapped `observations' $d_t-w_t$ of
demand-net-of-wind, including considerable positive autocorrelation
and shorter-term nonstationarity---the latter due to both diurnal and
seasonal effects.  Hence, in making the above uncertainty estimates,
the best that can reasonably be done is to block the data according to
historical season and to treat these season-long blocks as being
independent of each other.  Where, as described above, separate
estimates of LoLE and EEU are made based on each historical season of
data, then confidence intervals for the long-run LoLE and EEU are
given by the confidence intervals for the means of the seven
independent historical-season-based estimates of these
quantities. Since the individual season-based estimates of LoLE and
EEU are far from normally distributed over seasons, we use a
bootstrapping approach~\cite{Davison1997} in which, for example, the
seven historical-season based estimates of LoLE are sampled with
replacement to obtain a sufficiently large number of bootstrap
replications of the original set of seven estimates.  The distribution
of the means of these bootstrap datasets mirrors that of the overall
(sample) mean of the original seven season-based estimates, and so, in
the usual bootstrap approach, the quantiles of this distribution may
used to give confidence intervals for the `true' long-run LoLE.  These
confidence intervals for the long-run LoLE and, similarly, the
long-run EEU are arguably a little too narrow in that in each case the
bootstrap approach effectively treats the extremes of the seven
historical-season based estimates as representing the extremes of what
may happen in any given season.
% (usually the bootstrap approach is applied to larger datasets).
However, the purpose of this paper is to investigate the use of EVT
for estimating capacity adequacy metrics and---at least under the
assumption of independence between seasons---the above bootstrap
approach is sufficient to give a reasonably good approximation to the
sampling uncertainty associated with the long-run estimates of LoLE
and EEU.

As a comparison we also compute long-run LoLE and EEU estimates by
combining the seven historical seasons of (demand, wind) data and
obtaining pooled estimates of these quantities (i.e.\ using the full
seven-year dataset to compute long-run metrics rather than taking the
mean of the metrics corresponding to individual years). Confidence
intervals may still be obtained by using block
bootstrapping~\cite{Politis2003} in which the entire dataset is
re-sampled in season-long blocks (assumed independent) to obtain a
sufficiently large number of bootstrap replications of the entire
dataset.  Bootstrap estimates of LoLE and EEU are then obtained for
each of these replications, and confidence intervals obtained as
usual. Note that for the hindcast approach these two methods for
obtaining confidence intervals for the long-run LoLE and EEU estimates
will yield the same result. This is because the hindcast approach does
not smooth between years when data are pooled.

\section{Results}
\label{Results}

In this section the model described in Section \ref{Model} is fitted
to the GB data described in Section \ref{Data} and the LoLE and EEU
are estimated using this model.  The model is fitted using the
\texttt{ismev} package \cite{ismev} and the R computing
language~\cite{R}.

\subsection{Model fitting and validation}
\label{model-fitting}

\begin{figure}
  \centering
  \includegraphics[width=\linewidth]{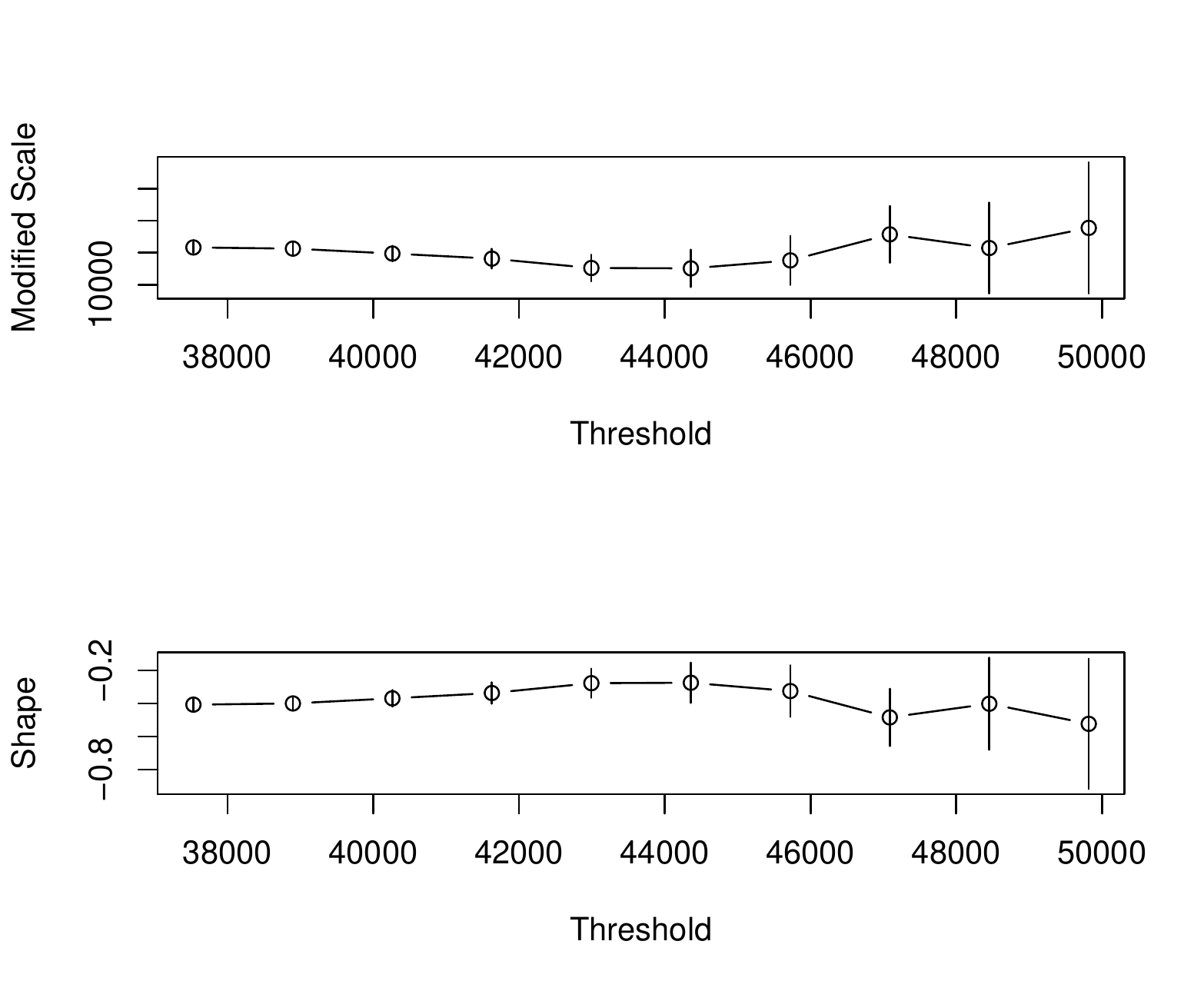}
  \caption{Estimated values of $\xi$ (bottom) and $\sigma^*$ (top) for
    different values of the threshold $u$ for the first season in the
    dataset.  The bars show the uncertainties in the
    estimates.
    \label{Threshold}}
\end{figure}

To fit the model \eqref{eq2} to the demand-net-of-wind data it is
first necessary to choose a threshold~$u$.  As results are to be
calculated by conditioning separately on each forward-mapped
historical season, different thresholds are used for each such season.
As described in Section \ref{Model}, the distribution of
demand-net-of-wind, for any given forward-mapped historical season, is
modelled by a generalised Pareto distribution above the threshold~$u$
and by its empirical distribution below the threshold~$u$.  The
threshold~$u$ must be sufficiently large that the required results
from extreme value theory hold, but a lower threshold means that more
data can be used for parameter estimation.  The aim is therefore to
use the lowest satisfactory threshold.  Following \cite{Coles2001} we
test a range of values for the threshold~$u$, fit the generalised
Pareto distribution \eqref{eq1} for each $u$ and assess the estimated
values of the shape parameter~$\xi$ and scale parameter~$\sigma_u$.
For this purpose, the latter is transformed to
$\sigma^*=\sigma_u-u\xi$ to remove the dependence of $\sigma_u$ on the
threshold~$u$: we require to choose a $u$ such that for all $u'>u$,
there is little variation in the estimated values of $\xi$ and
$\sigma^*$.

Figure \ref{Threshold} shows the estimated values of $\xi$ and
$\sigma^*$ for the first season of data, with thresholds~$u$ ranging
from around 38GW to 50GW (approximately the 99.5\% quantile).  The
uncertainties in the estimates of $\xi$ and $\sigma^*$ are clearly
increasing as the threshold~$u$ increases as less data above the
threshold is available for parameter estimation.  These uncertainty
estimates should be regarded as rough approximations, because their
calculation treats the data within a given season as consisting of
independent identically distributed observations, whereas, as
previously remarked, there is actually some serial correlation
structure within the data.  Nevertheless, a threshold~$u$ of around
45GW appears to be reasonable.  For thresholds less than 45GW there
seems to be a trend in both parameters.  For thresholds above 45GW the
uncertainty intervals overlap to such an extent that there is no
evidence to suggest that the parameter estimates are changing.  For
the first season in the dataset, 45GW corresponds approximately to the
95\% quantile of the forward-mapped demand-net-of-wind data for that
season.  Repeating the analysis described above for each of the later
seasons in the dataset leads to a similar conclusion---that in each
case the 95\% quantile is an appropriate choice of threshold.  To
further check the effect of threshold choice, the LoLE and EEU were
estimated using thresholds corresponding to the 90\%, 95\% and 98\%
quantiles of the forward-mapped demand-net-of-wind data associated
with each historical season.  These results are discussed later along
with further model validation to check that the fitted model is
consistent with the data.  The values of the 95\% thresholds used for
the analysis for each historical season of data, and also for a pooled
analysis which combines the data over historical seasons are given in
Table \ref{Parameters}.

\begin{table}[ht]
  \scriptsize
  \centering
  \begin{tabular}{|l|ccccccc|c|}
    \hline
    GW &07-08&08-09&09-10&10-11&11-12&12-13&13-14&All\\
    \hline
    95\% & 45.28 & 45.38 & 46.57 & 47.42 & 44.88 & 46.88 & 43.64 & 43.89 \\
    $\sigma_u$&2.85&2.57&2.07&2.92&2.51&2.38&2.57&2.51\\
    $\xi$&-0.32&-0.30&-0.22&-0.28&-0.24&-0.24&-0.38&-0.21\\
    \hline
  \end{tabular}
  \caption{Thresholds $u$ (GW) for each season in the dataset (the 95\%
    quantiles) and the estimated shape ($\xi$) and scale ($\sigma_u$)
    parameters corresponding to this 95\% threshold. Values are listed
    for individual seasons and for the full
    dataset.
    \label{Parameters}}
\end{table}

%thrv (middle threshold) 95 Q
%[1] 45277.81 45383.38 46573.03 47418.31 44880.02 46879.64 43638.21

%thrLv (lower threshold) 90 Q
%[1] 43407.51 43546.61 44526.41 45501.77 42249.75 44877.59 41880.62

%thrHv (upper threshold) 98 Q
%[1] 47477.64 47473.90 48314.34 49713.74 46920.37 48710.97 45522.88

%95 quantile for all data
%45907.99 

%90 quantile for all
%43885.42 

%98 quantile for all
%48067.69 

Given the thresholds in Table \ref{Parameters}, maximum likelihood can
be used to estimate the parameters $\xi$ and $\sigma_u$ in model
\eqref{eq1} (using the \texttt{ismev} package).  The parameter
estimates are given in Table \ref{Parameters} for the threshold
corresponding to the 95\% quantile of the distribution of
demand-net-of-wind.  As shown, the estimates of $\xi$ and $\sigma_u$
are reasonably consistent from season to season, although the
thresholds are variable. This year-to-year variability in the
threshold suggests that a pooled analysis may not be entirely
appropriate as datapoints that are extreme in one year may not be in
another.

\begin{figure}
  \centering
  \includegraphics[width=\linewidth]{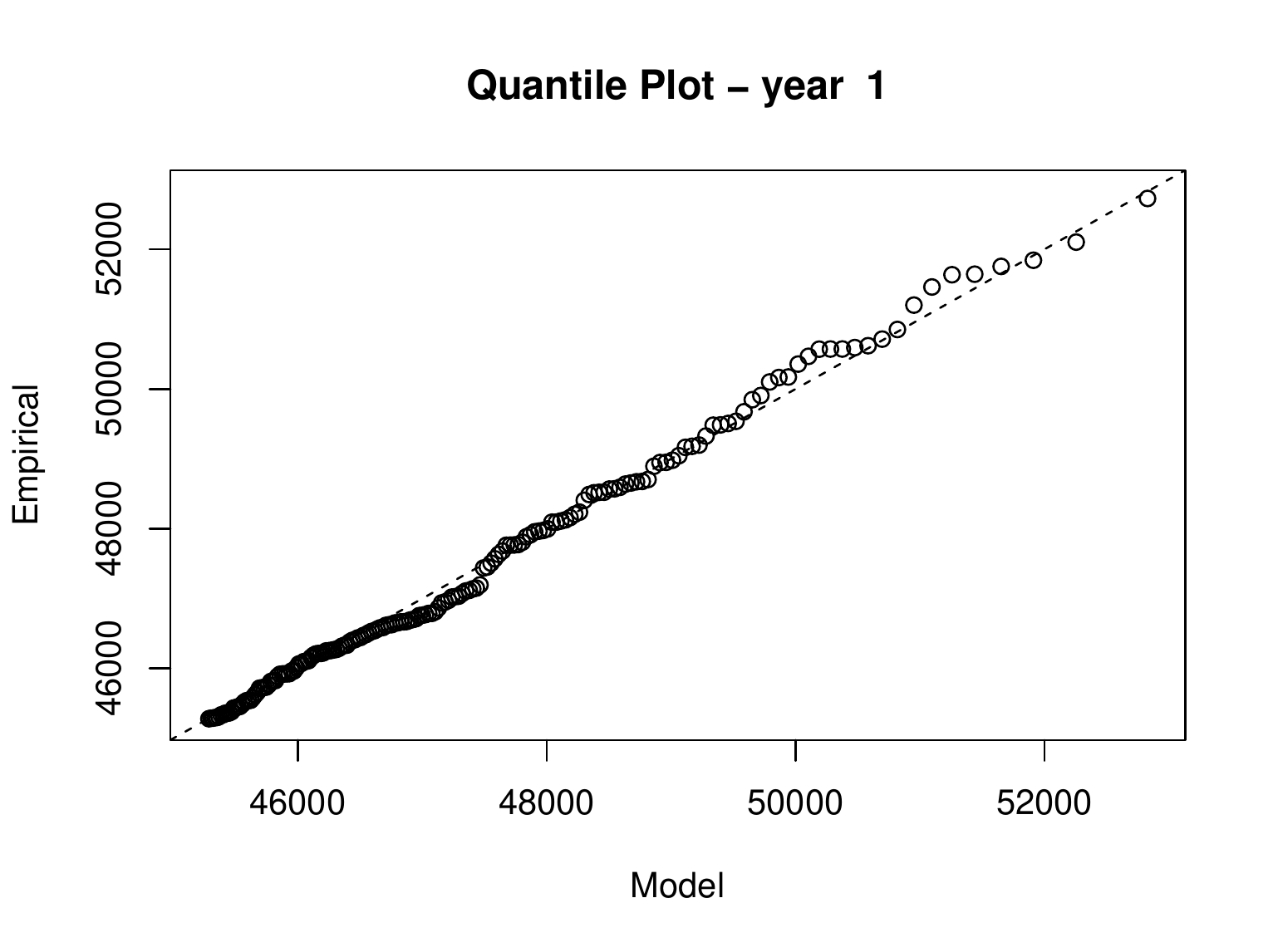}
  \caption{Quantile-quantile plot showing the fit of model \eqref{eq2}
    to the extreme tail of demand-net-of-wind for the first season in
    the dataset.
    \label{QQplot}}
\end{figure}

The fitted model can be validated by comparison to the observed data.
Figure \ref{QQplot} is a quantile-quantile plot of the tail of the
demand-net-of-wind data associated with the first historical season
(i.e.\ the demand-net-of-wind data over the EVT threshold for that
season) against the corresponding fitted model~\eqref{eq2}.  The
observed data are shown as circles.  If the data followed the model
\eqref{eq2} exactly they would lie on the dotted diagonal line shown.
As can be seen, the data do very closely follow the fitted model.
Quantile-quantile plots for the other historical seasons showed a
similarly good fit.

% Figure \ref{Tail} is a histogram of the extreme forward-mapped
% demand-net-of-%wind associated with the first season of the dataset.
% Overlaid is the fitted %generalised Pareto distribution.  The fitted
% curve appears to be consistent with the data, providing further
% evidence that the model is a good fit.  Similar results were
% obtained for later seasons. \sz{Does this not simply replicate (less
% satisfactorily) the info in the quantile plot?  If we keep it (which
% we shouldn't have to), we should
% make this clear.}

%\begin{figure}
%\centering
%\includegraphics[width=\linewidth]{Tail1.pdf}
%\caption{Histogram of the extreme tail of demand-net-of-wind of the
% first season in the dataset.  Smoothed line shows the fitted model
% \eqref{eq2}.  \label{Tail}}
%\end{figure}

\subsection{Estimation of LoLE and EEU}
\label{estimation-lole-eeu}

It follows from~\eqref{LoLE} and~\eqref{EEU} that estimation of LoLE
and EEU is based on estimation of the distribution of the
(non-sequential) supply-demand balance~$Z$ which is given
by~\eqref{sdbalance} and is the convolution of the corresponding
distributions of demand-net-of-wind $D-W$ and conventional
generation~$X$.  The fitted distribution of $D-W$ associated with any
forward-mapped historical season of (demand, wind) data is entirely
described by the tail function given in \eqref{eq2} above the
threshold~$u$ and by the empirical distribution of the associated
observations $d_t-w_t$ below that threshold.  The distribution of
conventional generation~$X$ is formed as described in
\cite{Billinton2012}.  Estimates of the capacities and availability
probabilities of the conventional generators on the system in the
future season of interest were obtained from National Grid.  Random
errors were added to the capacities to protect the sensitivity of the
data.  As such, the results presented should be seen as broadly
representative of the GB system but do not provide accurate estimates
of the risk in that system.  Each generator is assumed to provide full
capacity with its availability probability and otherwise to provide
zero capacity.  The generators are assumed to be independently
available, and so the convolution of their individual two-state
distributions gives the distribution of the total available
conventional generation~$X$.  The distribution of~$Z$---and so also
estimates of LoLE and EEU---for the future season under study, based
on any given historical season of (demand, wind) data, are then
obtained as described above.  The convolution of the distribution of
$X$ with that of $D-W$ is obtained by the discretisation of the latter
in 1~MW bins.

Estimates of LoLE and EEU conditional on each historical season of
(demand, wind) data, and for the three different choices of
threshold~$u$ as above, are given in Tables \ref{LoLEresults} and
\ref{EEUresults}.  The results are broadly similar for all three
thresholds, suggesting that these estimates are not sensitive to the
precise choice of threshold.

In Tables \ref{LoLEresults} and \ref{EEUresults} these estimates can
be compared to those obtained using two alternative approaches:
hindcast and a model which assumes independence between demand and
wind.  The hindcast approach estimates the probability $P(D-W>v)$ by
the empirical proportion of the observations $d_t-w_t$ which are
greater than $v$. The model assuming independence fits separate
empirical distributions to the demand and wind data for each
historical season. The distributions for wind and demand are then
convoluted to obtain the distribution of $D-W$. As shown in Tables
\ref{LoLEresults} and \ref{EEUresults}, the results obtained using the
hindcast approach are similar to those obtained using EVT, especially
to those EVT results obtained using a 98\% threshold.  (The latter
observation is unsurprising since, as the threshold~$u$ is increased,
the EVT analysis becomes closer to the hindcast.)  However, using a
lower threshold provides a greater degree of smoothing, inferring more
information from less extreme data, and thereby providing results
which are more robust to small changes in the data.  The results
obtained using the assumption of wind-demand independence are similar
to those obtained using the EVT and hindcast approaches for some
seasons of historical data and significantly different for others
(e.g.\ 2008--09, 2009--10, 2012--13).  For the 2008--09 and 2009--10
historical data the risk levels obtained using the independence
assumption are higher, while for the 2012--13 historical data the risk
level obtained using the independence assumption is lower.  Since it
is the above independence assumption which is suspect here, these
results highlight the dangers involved in making it.

%i.e.\ for each historical season,
%\begin{equation}
%P(D-W > v)= \frac{1}{n} \sum_{t=1}^n I(d_t-w_t>v), \label{hindcast}
%\end{equation}
%where $I(\cdot)$ is an indicator function taking the value of $1$ if $d_t-w_t>v$ and %$0$ if $d_t-w_t\leq v$.  

%\begin{equation}
%P(W > w)= \frac{1}{n} \sum_{t=1}^n I(w_t>w), \label{IndWind}
%\end{equation}
%and
%\begin{equation}
%P(D> d)= \frac{1}{n} \sum_{t=1}^n I(d_t>d).  \label{IndDemand}
%\end{equation}

%% LOLE results

\begin{table}[ht]
  \scriptsize
  \centering
  \begin{tabular}{|l|ccccc|}
    \hline
    Season & EVT 90\% & EVT 95\% & EVT 98\% & Hindcast & Ind\\ 
    \hline
    07-08 & 2.86 & 2.82 & 3.00 & 3.07 & 2.74 \\ 
    08-09 & 2.21 & 2.22 & 2.25 & 2.29 & 3.05 \\ 
    09-10 & 4.43 & 4.02 & 3.90 & 3.85 & 5.45 \\ 
    10-11 & 16.33 & 16.77 & 17.63 & 17.60 & 17.81 \\ 
    11-12 & 2.17 & 1.92 & 1.92 & 1.95 & 1.56 \\ 
    12-13 & 7.87 & 7.69 & 7.57 & 7.97 & 5.64 \\ 
    13-14 & 0.16 & 0.15 & 0.15 & 0.17 & 0.26 \\ 
    \hline
    Mean&5.15&5.08&5.20&5.27&5.22\\
    CI&(2.02,9.25)&(1.92,9.37)&(1.95,9.71)&(1.97,9.79)&(2.01,9.70)\\
    \hline
  \end{tabular}
  \caption{LoLE (hours per season) conditional on the demand-wind
    profile of each season in the dataset, estimated using: extreme
    value theory (with thresholds of the 90\% quantile, 95\% quantile
    and 98\% quantile), hindcast and the independence model (Ind).
    Confidence intervals account for sampling uncertainty in demand
    and wind.
    \label{LoLEresults}}
\end{table}
% CIs calculated using t dist give negative lower bounds.

%% EEU results

\begin{table}[ht]
  \scriptsize
  \centering
  \begin{tabular}{|l|ccccc|}
    \hline
    Season & EVT 90\% & EVT 95\% & EVT 98\% & Hindcast & Ind\\ 
    \hline
    07-08 & 2.99 & 2.81 & 2.92 & 3.03 & 3.02 \\ 
    08-09 & 2.14 & 2.12 & 2.13 & 2.18 & 3.07 \\ 
    09-10 & 4.56 & 4.15 & 4.07 & 4.09 & 6.11 \\ 
    10-11 & 24.07 & 24.01 & 25.01 & 25.92 & 24.77 \\ 
    11-12 & 2.11 & 1.95 & 1.96 & 2.05 & 1.45 \\ 
    12-13 & 9.22 & 9.16 & 9.16 & 9.73 & 6.17 \\ 
    13-14 & 0.11 & 0.10 & 0.10 & 0.12 & 0.19 \\ 
    \hline
    Mean&6.46&6.33&6.48&6.73&6.40\\
    CI&(2.02,12.73)&(1.91,12.61)&(1.92,13.04)&(1.97,13.53)&(2.07,12.84)\\
    \hline
  \end{tabular}
  \caption{EEU (GWh per season) conditional on the demand-wind profile
    of each season in the dataset, estimated using: extreme value
    theory (with thresholds of the 90\% quantile, 95\% quantile and
    98\% quantile), hindcast and the independence model (Ind).
    Confidence intervals account for sampling uncertainty in demand
    and wind.
    \label{EEUresults}}
\end{table}

The results in Tables \ref{LoLEresults} and \ref{EEUresults} show
substantial variability between seasons.  The LoLE ranges from around
0.15 to 16.77 hy$^{-1}$ and the EEU ranges from around 0.1 to 24.01
GWhy$^{-1}$.  That these ranges are wide highlights the need to
consider the variability of the risk level with weather conditions.
Decision-makers might be very averse to an LoLE or EEU at the higher
end of this range but happy with the overall mean.  Note that the
estimates of LoLE and EEU conditional on a given demand-net-of-wind
profile still integrate over uncertainty in conventional generation
(and hence are still expected values).  The actual variation in
loss-of-load and energy unserved from season to season will therefore
be larger than the variation shown between seasons in Tables
\ref{LoLEresults} and \ref{EEUresults}.

Tables \ref{LoLEresults} and \ref{EEUresults} also give estimates of
the long-run LoLE and long-run EEU.  These are the means of the
estimates based on the individual historical seasons of
demand-net-of-wind data.  The 95\% confidence intervals for these
long-run estimates are calculated via bootstrapping as described in
Section~\ref{Model}, i.e.\ by regarding the estimates for the seven
historical seasons as seven independent observations. The confidence
intervals for the long run estimates therefore reflect uncertainty
arising from the limited number of seasons of data (see also
Section~\ref{Modelb}).  These confidence intervals are wide, ranging
from around 2 to 10 hy$^{-1}$ for LoLE and from 2 to 13 GWhy$^{-1}$
for EEU, reflecting the considerable variability in the estimates
based on individual seasons of data.  Again, this variability
demonstrates the importance to decision-makers of understanding this
uncertainty.  The widths of the confidence intervals increase slightly
with increasing EVT threshold and are greatest for those based on the
hindcast approach.  This is to be expected---the smoothing provided by
the EVT approach reduces variability because more data is used to
estimate the far right tail of the supply-demand balance.  The widths
of the confidence intervals obtained under the assumption of
demand-wind independence are comparable to those obtained using EVT
but, as described above, the use of the independence assumption risks
biasing the estimates of LoLE and EEU.

Table \ref{Results_all} gives pooled estimates of LoLE and EEU
obtained by fitting the above models to all seven historical seasons
of data simultaneously (in contrast to fitting the models to each
season individually).  The EVT thresholds used are the corresponding
quantiles of the demand-net-of-wind distribution for the full dataset.
% , rather than using a different threshold for each historical season.
For the EVT and hindcast approaches, the long-run LoLE and EEU
estimates are similar to those already obtained as the means of the
individual season-based estimates and reported in Tables
\ref{LoLEresults} and \ref{EEUresults}.  The corresponding 95\%
confidence intervals---obtained using block bootstrapping with
season-long blocks---are unsurprisingly also similar to those already
obtained, with the hindcast approach again giving the widest
confidence intervals.  For the model based on the assumption of
demand-wind independence, the pooled estimates of LoLE and EEU are
respectively 4.46 hy$^{-1}$ and 5.30 GWhy$^{-1}$, in contrast to the earlier
individual season-based estimates of 5.22 hy$^{-1}$ and 6.40 GWhy$^{-1}$.  For
this independence model the confidence intervals obtained via the
pooled analysis are smaller than those obtained previously.  These
results suggest that the assumption of demand-wind independence may be
more problematic when applied across multiple seasons of data.  One
possible reason is that long-term weather regimes may induce
dependence between demand and wind generation aggregated over multiple
seasons, while this dependence may largely disappear when conditioning
on individual seasons (as in Tables \ref{LoLEresults} and
\ref{EEUresults}).  These results suggest that if a demand-wind
independence model is used, better results may be obtained by fitting
the model separately to each season in the dataset and then obtaining
risk metrics by averaging over these seasons.

% Results when all seasons combined
\begin{table}[ht]
  \scriptsize
  \centering
  \begin{tabular}{|l|ccccc|}
    \hline
    & EVT - 90\%& EVT - 95\% & EVT - 98\% & Hindcast & Ind \\ 
    \hline
    LoLE & 5.36 & 5.26 & 5.10 & 5.27 & 4.46 \\ 
    95\% CI&(1.96,9.46)&(1.93,9.33)&(1.93,9.52)&(1.97,9.79)&(1.79,8.66)\\
    EEU& 6.53 & 6.52 & 6.55 & 6.73 & 5.30 \\ 
    95\% CI&(1.94,12.72)&(1.95,12.84)&(1.95,12.92)&(1.97,13.53)&(1.84,11.28)\\
    \hline
  \end{tabular}
  \caption{LoLE (hours per season) and EEU (GWh per season) estimated
    by fitting the models to the full seven season forward-mapped
    demand-net-of-wind dataset.  Models used: extreme value theory
    (with threshold of the 90\%, 95\% and 98\% quantiles), hindcast
    and the independence model (Ind).
    \label{Results_all}}
\end{table}

\section{Conclusion}

This paper has investigated the use of extreme value theory (EVT) for
modelling the distribution of demand-net-of-wind for capacity adequacy
assessment.  The main advantage of this approach is that EVT provides
a mathematically-justified mechanism for estimating the extreme right
tail of the distribution of demand-net-of-wind (corresponding to times
of high demand and low wind); this is normally the only part of the
distribution which is relevant for capacity adequacy.  This smoothing
involved in this estimation reduces the effect of outliers and small
variations in the tail data when compared to use of the empirical
distribution.  A further advantage of this approach is that
observations of demand-net-of-wind can be used directly, meaning that
there is no need to make strong parametric assumptions about the
underlying distributions of the demand and wind processes, or about
the nature of the dependence between demand and wind.

The paper has also shown that typically estimates of risk metrics such
as LoLE and EEU vary greatly according to the historical winter season
of (demand, wind) data used in the estimation process, indicating a
strong dependence on the prevailing weather in the winter season under
study.  This has two consequences: first, actual outcomes for these
metrics in any given future season may be very different from
estimated long-run averages; second, uncertainty estimation for these
long-run averages can probably only be satisfactorily made by blocking
data according to historical season and treating (demand, wind)
regimes in distinct winter seasons as independent---observations
within seasons may not be treated as independent of each other. The
first consequence is further compounded because there are usually only
a small number of relevant years of data for the estimation of risk
metrics (seven in our example), meaning that it is unlikely that the
full year-to-year variability has been captured in the dataset.

\section*{Acknowledgment}

The authors would like to thank Iain Staffell for providing the wind
and demand data and National Grid for providing data on conventional
generation.  They are further grateful to Chris Dent and to colleagues
at National Grid for helpful comments and discussions.

\bibliographystyle{IEEEtran}
\bibliography{bib}

\vskip -2\baselineskip plus -1fil

\begin{IEEEbiographynophoto}{Amy L.  Wilson}
  is a Research Associate in the School of Mathematics, University of
  Edinburgh, U.K.  She obtained her MMath degree in Mathematics from
  the University of Cambridge, U.K. and her PhD in Statistics from the
  University of Edinburgh.  Her current research interests include the
  analysis of risk for capacity adequacy studies and quantification of
  uncertainty in large scale energy system models.
\end{IEEEbiographynophoto}

\vskip -2\baselineskip plus -1fil

\begin{IEEEbiographynophoto}{Stan Zachary}
  is a Senior Researcher in the School of Mathematics, University of
  Edinburgh, U.K, and an Honorary Fellow of Heriot-Watt University,
  Edinburgh.  He obtained his MMath degree in Mathematics from
  the University of Cambridge, U.K. and his PhD in Statistics from the
  University of Durham.  His current research interests include the
  probabilistic, statistical and economic modelling of large energy
  systems.
\end{IEEEbiographynophoto}

\end{document}